\documentclass[12pt]{article}
\usepackage{color} %delete this after comments are gone - Aleena
\usepackage{graphicx} % Include figure files, delete for final version -- Aleena
\usepackage{ragged2e} % Include figure files, delete for final version

\usepackage{scicite}
\usepackage{times}

\newenvironment{sciabstract}{%
\begin{quote} \bf}
{\end{quote}}

\title{Observation of liquid glass in suspensions of ellipsoidal colloids}

\author
{J\"org Roller,$^{1}$ Aleena Laganapan,$^{2}$ Janne-Mieke Meijer,$^{2,3}$ \\
Matthias Fuchs,$^{2\dagger}$ Andreas Zumbusch$^{1\ast}$\\
\\
\normalsize{$^{1}$University of Konstanz, Dep. of Chemistry, Germany}\\
\normalsize{$^{2}$University of Konstanz, Dep. of Physics, Germany}\\
\normalsize{$^{3}$University of Amsterdam, Institute of Physics, The Netherlands}\\
\\
\normalsize{To whom correspondence should be addressed;} 
\\
\normalsize{E-mail:  $^\dagger$matthias.fuchs@uni-konstanz.de, $^\ast$andreas.zumbusch@uni-konstanz.de}
}

\date{}

\newcommand{\rem}[1]{\textcolor{green}{}}

\begin{document} 


%\baselineskip24pt


\maketitle


\begin{sciabstract}
Despite the omnipresence of colloidal suspensions, little is known about the influence of shape on phase transformations, especially in nonequilibrium. To date, real-space imaging results are limited to systems composed of spherical colloids. In most natural and technical systems, however, particles are non-spherical and their structural dynamics are determined by translational and rotational degrees of freedom. Using confocal microscopy, we reveal that suspensions of ellipsoidal colloids form an unexpected state of matter, a liquid glass in which rotations are frozen while translations remain fluid. Image analysis unveils hitherto unknown nematic precursors as characteristic structural elements of this state. The mutual obstruction of these ramified clusters prevents liquid crystalline order. Our results give unique insight into the interplay between local structures and phase transformations. This helps to guide applications such as self-assembly of colloidal superstructures and also gives first evidence of the importance of shape on the glass transition in general.
\end{sciabstract}

\section*{Introduction}
Suspensions of colloidal particles are widely spread in nature and technology and have been studied intensely over more than a century. When the density of such suspensions is increased to high volume fractions, often their structural dynamics are arrested in a disordered, glassy state before they can form an ordered structure. This puts limits on technical applications such as the formation of superstructures from colloidal particles by self-assembly processes. Yet, it also means that knowledge gained from investigations of phase transformations of colloidal suspensions to ordered structures and glasses provides insight into similar 
phenomena in a broad range of complex glass forming materials, ranging from metals to biological cells \cite{Angelini2011, Parry2014, Pueblo2017}. 

The colloidal glass transition has therefore been considered a model that features many of the glass transition phenomena found in atomic and molecular systems \cite{Weeks2017}. Being big enough to allow their real-space observation using optical microscopy, but small enough to remain suspended over extended periods, colloids in such model systems are employed as ‘big atoms’ \cite{Poon2004} which can be investigated on an individual particle basis \cite{Hunter2012, Gokhale2016}.  
By analyzing the trajectories of thousands of particles, detailed insights into glass phenomena, such as dynamic heterogeneities of collectively rearranging structures have been obtained \cite{Kegel2000, 
Weeks2000}. Apart from their use as model systems, however, synthetic colloids have increasingly also been perceived as interesting material building blocks in their own right \cite{Manoharan2015, Li2016}. The recent growth of this field of research has been supported by a multitude of novel techniques for the synthesis of colloidal particles with specific geometries and interactions \cite{Glotzer2007, Sacanna2011, Porter2017}. 

The availability of shape-anisotropic particles allows investigations of their phase transformations in dense suspensions which promises to give unique insight into structural dynamics of complex systems. This is especially important for the investigation of steric effects, which need to be controlled in the self-assembly of colloidal building blocks into materials with specific collective properties. The simplest deviation from spherical symmetry is uniaxial stretching to prolate ellipsoids \cite{Keville1991}. Already for these simple particles, theory and simulations predict a rich phase diagram \cite{Odriozola2012} and complex glass formation \cite{Letz2000, DeMichele2007, Pfleiderer2008, Chong2005} due to the presence of translational and rotational degrees of freedom. However, the few investigations of suspensions of ellipsoids have all been focused on static structures obtained by driving particles in external gravitational or electric fields \cite{Mohraz2005, Crassous2014, Ganesan2017, Pal2018}. The only investigations on steric effects of ellipsoid shape have been performed in 2D films \cite{Zheng2014, Mishra2014}. To date, hardly any experimental data on the influence of steric factors on the phase transformation in 3D ellipsoidal suspensions exist.  

Here, we present the first particle-resolved studies of the structural dynamics of ellipsoidal colloid suspensions. The experiments were performed on a large range of different volume fractions. Mode coupling theory (MCT) predicts that in systems of this type, a liquid glass should exist in which particle rotation is frozen whereas their translation is still liquid \cite{Letz2000}. We used quantitative optical microscopy in combination with a novel type of core-shell particles \cite{Klein2014} to simultaneously track translational and rotational particle motion with high precision \cite{Roller2018}. Upon increasing the volume fraction $\phi$, we find that rotational degrees of freedom undergo a glass transition before translational dynamics are arrested such that a liquid glass is
formed. To understand the nature of the observed rotational and translational glass transitions, we further corroborate our results with an MCT analysis and Brownian dynamics simulations and find that long-range correlations are the cause for the emergence of this unique state. Detailed image analysis reveals that hitherto unknown nematic precursors exist as the characteristic structural elements of the liquid glass. They consist of ramified aligned regions intersected by differently ordered or disordered regions and appear to mediate the long-range correlations present in the liquid glass state of particle suspensions. Our experiments give a first impression of the complex behavior of colloidal suspensions arising from the introduction of the simplest geometrical distortion of the particles' shape from sphericity. They show how sterical factors lead to the emergence of peculiar local structures mediating long range spatial correlations which result in the formation of amorphous states preempting the globally ordered state. % as a new state of matter.  

%%%%%%%%%%%%%%%%%%%%%%%% END OF INTRODUCTION%%%%%%%%%%%%%%%%%%%%%%%%%%%%%

\section*{Results and Discussion}

In our experiments, we used ellipsoidal polymethylmethacrylate (PMMA) colloids with a long semi-axis of $a = 4.32\,\mu$m and a 
short semi-axis of $b = 1.23\,\mu$m, i.e. an aspect ratio of $a/b = 3.5$ (Fig. 1A). The particles were sterically 
stabilized and suspended in a density- and refractive-index-matched solvent mixture \cite{Royall2012}. Tracking 
of particle positions and orientations was facilitated by using particles with core-shell geometry where the spherical core and the ellipsoidal shell were labeled with different fluorphores \cite{Klein2015}. Using confocal laser scanning microscopy, we typically recorded the temporal development of the 3D particle positions and orientations for more than 6000 particles with accuracies of $60\,$nm and $5^\circ$, respectively \cite{Roller2018} (Fig. 1B). Since volume fractions $\phi$ are the pivotal variable in all the experiments, we took utmost care in their determination (see Methods for details). In order to validate the experimental results, we performed event-driven Brownian dynamics (ED-BD) simulations that model overdamped dynamics of hard ellipsoids \cite{Scala2007}. The system consisted of $N$ non-overlapping ellipsoids in a cubic simulation box of length $L$ that was varied depending on the desired $\phi$. Results from experiments and computer simulations were analyzed using MCT.  

\begin{figure}[htp]
\centering
\includegraphics[width=0.7\columnwidth]{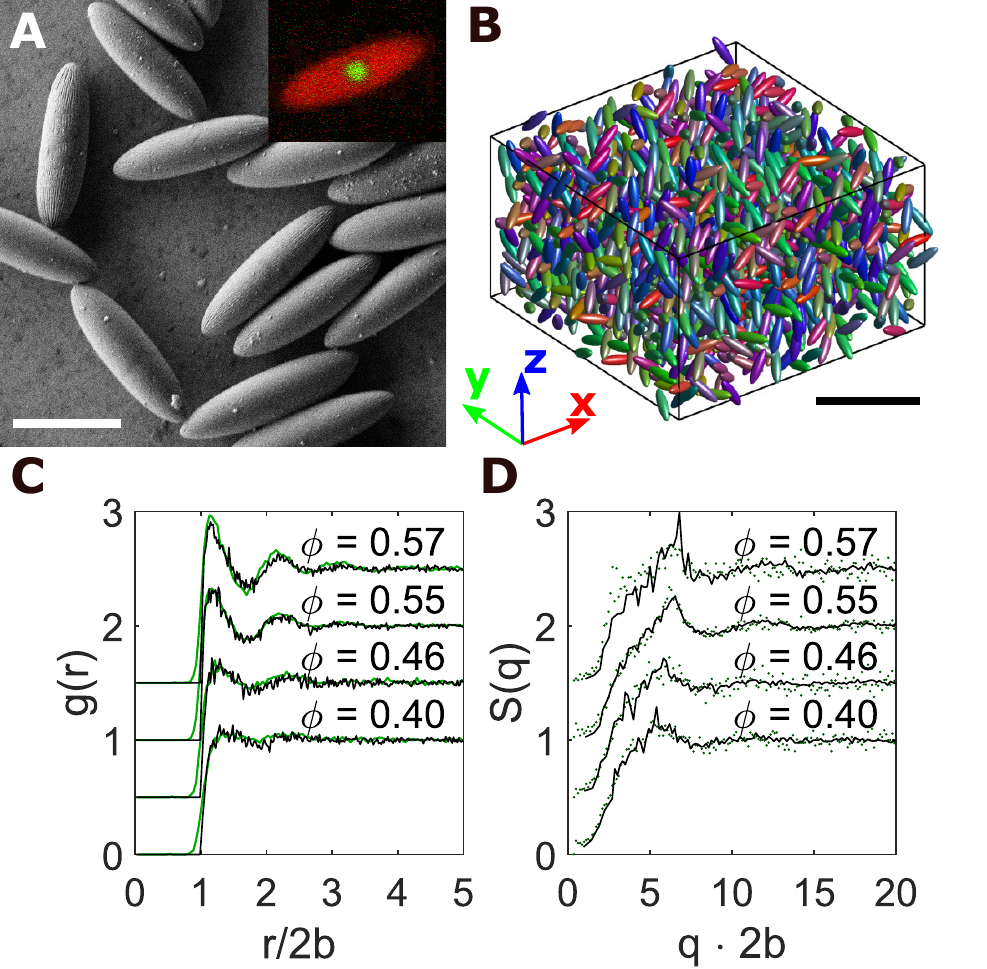}
\justify{\textbf{Fig.~1. Characterization of ellipsoidal colloids and their packings: } Scanning electron microscope image (\textbf{A}) of the ellipsoidal colloids with aspect ratio $a/b=3.5$. The inset shows a confocal microscope image, highlighting the core-shell structure. Scale bar is $5\,\mu$m. (\textbf{B}) Computer rendered 3D reconstruction of a subset of a sample volume at $\phi=0.57$ with the RGB value of the color indicating the particle orientations. Scale bar is $20\,\mu$m. (\textbf{C}) Pair correlations function $g(r)$ and (\textbf{D}) structure factor $S(q)$ for $\phi$ as labeled from experiment (green points) and simulation data (black lines). Distances $r$ are rescaled by the ellipsoid width $2b$.}
\end{figure}

\subsection*{Structural correlations and lack of order}

To gain insight into the systems' behavior, we first extracted the pair correlation functions $g(r)$ and equilibrium structure factors $S(q)$ from the 3D particle positions at different $\phi$ (Fig. 1C,D). With increasing $\phi$, we observed the emergence of next-neighbor and second next-neighbor peaks in $g(r)$ while long range correlations were absent. At the same time in $S(q)$, the inverse peak position $2\pi/q_{\rm max}$, indicating the average particle separation, became smaller with increasing density. All observations are typical for liquid-like structural correlations and show that the systems remained in a disordered state without translational order for all investigated $\phi$. Fig. 1 also reveals the excellent coincidence between the measured $g(r)$ and $S(q)$ with those obtained by the ED-BD simulations. This confirms that in the suspensions, our elliptical colloids interact like hard particles. 

\begin{figure*}[h!]
\centering
\includegraphics[width=\columnwidth]{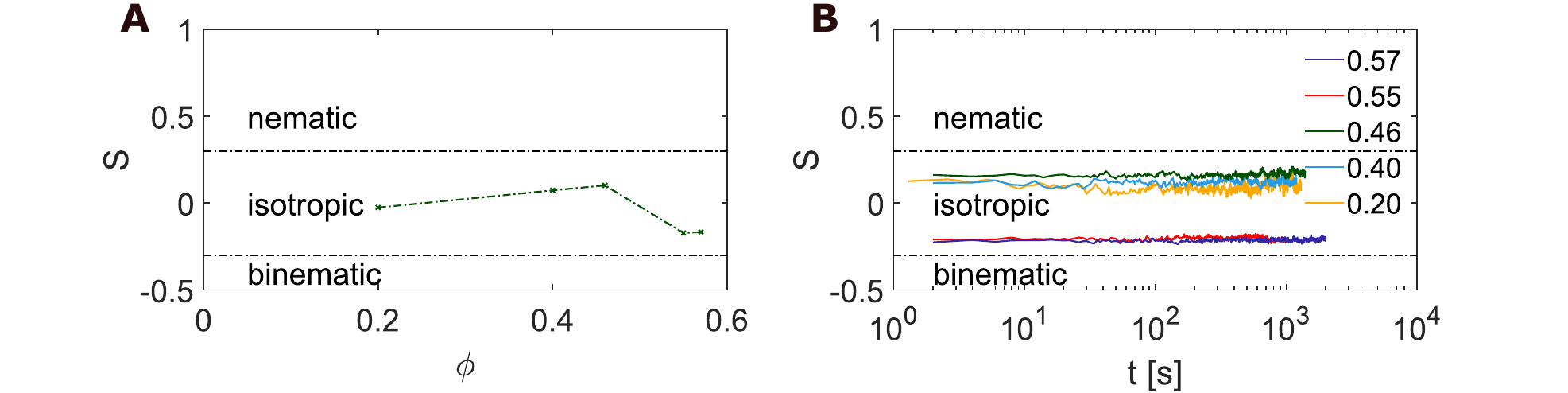}
\justify{
\textbf{Fig. 2. Nematic order parameter $S$ obtained from the nematic tensor $\mathbf{Q_{ij}}$.}  Panel (\textbf{A}) shows $S$ for different $\phi$, indicating the absence of a global particle alignment in the system. (\textbf{B}) Time evolution of the nematic order parameter for different $\phi$, again showing no signs of an evolving directional order. }
\end{figure*}

The equilibrium phase diagram of ellipsoids predicts the transition to a nematic state with increasing $\phi$ \cite{Odriozola2012}, thus we determined the orientational order of the whole sample. It  was probed by calculating the nematic order parameter $S$ \cite{DeMichele2007}, which is the eigenvalue possessing the largest absolute value of the nematic order tensor $Q_{ij}$. The nematic order tensor is given by
\begin{equation}
Q_{ij} = \left\langle \frac{3}{2} u_iu_j - \frac{1}{2} \delta_{ij}\right\rangle,\label{eq:Qij}
\end{equation}
where $\vec{u}$ is an eigenvector representing the orientation of the ellipsoid. The order parameter $S$ for different measured $\phi$ shows no indications for nematic order in the probed systems (Fig.~2A). For  all $\phi$, it remains below a value of $S < 0.3$ which is the commonly accepted criterion for the isotropic-nematic transition in simulations \cite{DeMichele2007}. The absence of nematic order is confirmed by plotting the changes of the order parameter $S$ with time (Fig.~2B) for different $\phi$. It shows no temporal evolution for all $\phi$, indicating the stability of the observed isotropic states.

\subsection*{Dynamical signatures of glass formation}

The absence of structural order even in very dense suspensions hints at the possible formation of a glass. To verify this, it is necessary to investigate the structural relaxation dynamics of the suspensions. The translational and rotational dynamics of the system are contained in the temporal evolution of the self part of the density correlation function
\begin{equation}
F_s(q,t) = \frac 1N
\sum_{i}^N  \langle \exp \left[i \vec{q}\cdot(\vec{r}_i(t' + t) - \vec{r}_i(t'))\right]\rangle
\end{equation}
and of the orientational correlation functions
\begin{equation}
L_n(t)= \frac 1N \sum_i^N \langle P_n(\cos(\theta_i(t' +t)-
\theta_i(t')) \rangle,
\end{equation}
respectively. Here, $\vec r_i$ is the position of the center of particle $i$, $\theta_i$ its orientation relative to a fixed laboratory direction, and $N$ the total particle number. The $P_n$ are the Legendre polynomials of order $n$, and $\langle \rangle$ denotes averaging over $t'$. For the calculation of $F_s(q,t)$ and $L_n(t)$, the wave vector $q$ was set to different values $5.18< q\cdot2b<7.90$ and $n=2,4$ was chosen, respectively.

The left panels of Fig. 3
describe the translation behavior and depict the experimentally determined $F_s(q,t)$ values for each $\phi$. In panel A, the wavevector is set to $q=2.6\,\mu$m$^{-1}$ (viz.~$2bq=6.42$) which corresponds to the maximum in the static structure factor and thus to the average particle distance. In panel C, a larger wavevector, 
$q=3.2\,\mu$m$^{-1}$ ($2bq=7.90$) is selected, where the relaxation is faster and the final decay can be resolved well in fluid states. Correlation functions obtained for different values of $q$ can be superimposed by scaling with $\log(tq^2)$ for low $\phi$ as is expected for the dynamics of isolated ellipsoids (see Supplementary text). For a liquid state, one expects the translation correlators $F_s(q,t)$ to decay completely over time. During the measurement time, this was observed for volume fractions up to $\phi = 0.46$. The curves obtained for $\phi = 0.55$ are already stretched and thus show a slowing down of the decay dynamics, but one can still assume their complete decay albeit at times longer than those probed. States at $\phi \le 0.55$ thus are also fluid-like. By contrast, at the highest volume fraction $\phi = 0.57$ the translational correlator does not decay anymore which is see by the clear plateau for translational motion typical for a glass state appears. From these observations we conclude that for translational dynamics, a glass transition occurred between $\phi=0.55$ and $\phi=0.57$. 
 
\begin{figure}[htp]
\centering
\includegraphics[width=\columnwidth]{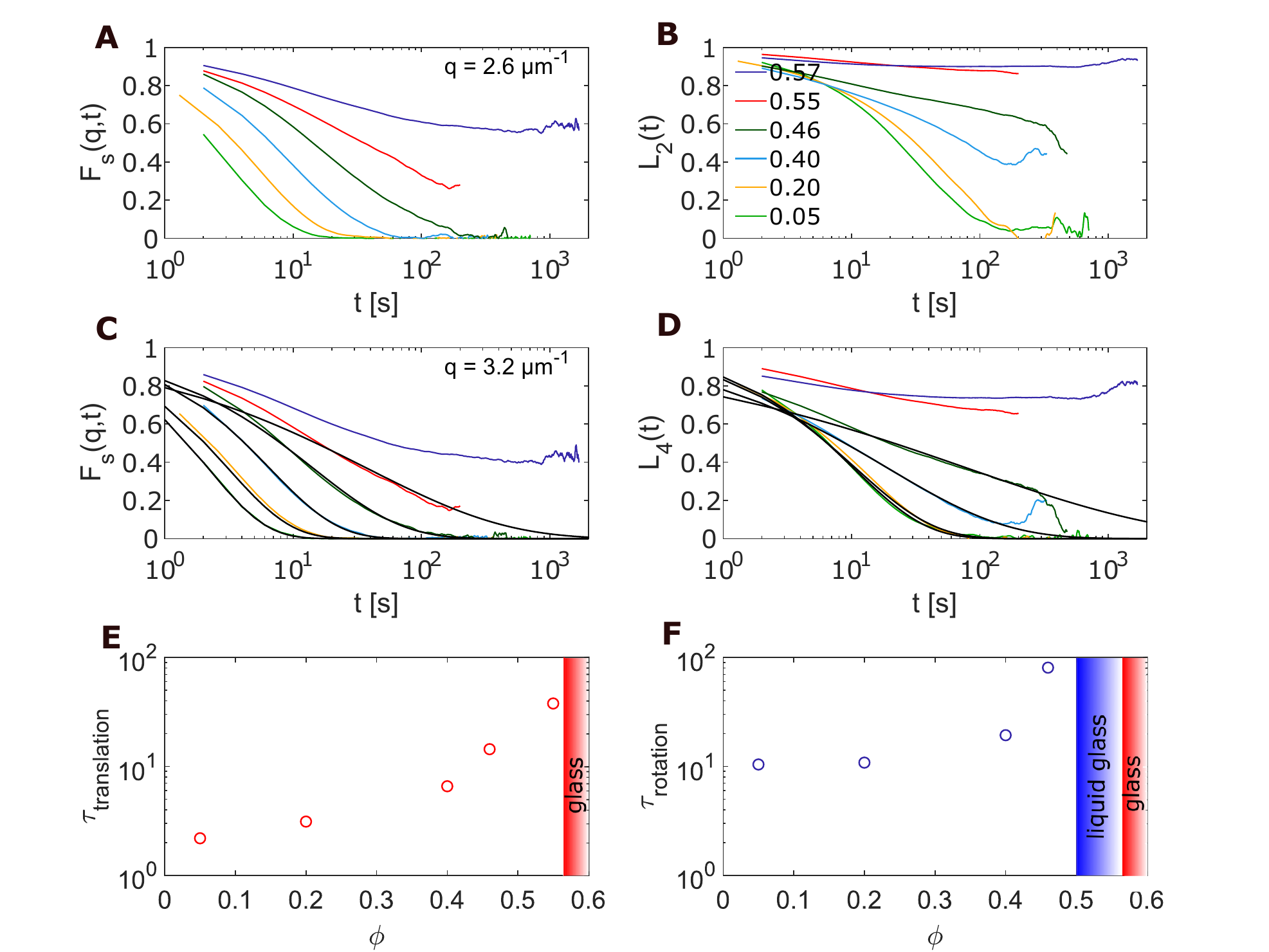}
\justify{\textbf{Fig. 3. Temporal correlation functions capturing translational and rotational dynamics  for different $\mathbf{\phi}$.} (\textbf{A}) and (\textbf{C}) depict the self part of the density correlation $F_s(q,t)$ at wavelengths comparable to the average particle separation, viz.~$2bq=6.42$ (A) and  $2bq=7.90$ (B), respectively. (\textbf{B}) and (\textbf{D}) show  the orientational correlation function $L_n(t)$ for $n=2$  and $n=4$; the legend in panel B gives the packing fractions for panels A-D.  Close to the glass transition, the decay of the curves is too slow to be captured within the measurement times. (\textbf{C}) and (\textbf{D}) include fits to the correlators in fluid states using  the Kohlrausch function Eq.~(\ref{eq:KWW}). (\textbf{E}) and (\textbf{F}) show the relaxation times obtained from the fits. The colored regions mark  the glass transitions $\phi_c^t$ and $\phi_c^{r}$ which were obtained by the MCT glass stability analysis (see Materials and Methods).}
\end{figure}

Fig. 3 B and D depict the temporal relaxation behavior of the 
rotation correlators $L_n(t)$ for $n=2$ and $n=4$, respectively. A comparison with their translational counterparts illustrated in Fig. 3 A and C, clearly shows the differences in the translational and rotational relaxation of the suspensions. For volume fractions $\phi \leq 0.46$, a complete relaxation of orientational correlation is observed even if the curves for $\phi = 0.46$ are already stretched significantly. However, for the two highest $\phi$ a clear plateau indicative of incomplete relaxation appears in both orientational correlation functions. These frozen orientational correlations are clear signs of glass-like behavior. Plateaus in the orientational correlations functions could also arise from nematic ordering. The observed values for the plateau heights, however, are too large to be compatible with the negligible nematic order of our samples. In nematic states, the relation $L_2(t\to\infty) = S^2$ holds \cite{Eppenga1984}, and the small values of the order parameter $S$ from Fig.~(2) are incompatible with the high amplitudes of frozen-in orientational correlations. We thus conclude that the data show  a glass transition in the rotational dynamics occurring between $\phi=0.46$ and $\phi=0.55$. 

The inspection of Fig.~3 leads to the conclusion that two different glass transitions, one in the orientational and a second one in the translational motion, exist. To gain more insight into this phenomenon, we turn to a quantitative analysis. In fully relaxing systems, the long-time decay, often termed $\alpha$-relaxation, can be fitted with a Kohlrausch function \cite{Stillinger2013}:
\begin{equation} 
\Phi(t) = f_{\Phi} \exp(-(t/\tau)^{\beta}), \label{eq:KWW}
\end{equation} 
where $f_{\Phi}$ is an amplitude which was set to $f_{\Phi}=1$, $\tau$ is the relaxation time, and $\beta$ is the so-called stretching exponent. Using this function we obtain fits that agree very well with our data for curves showing a clear decay within the experimental time window (Fig. 3 C and D; fit parameters are collected in the Supplementary text). Results for the translational and rotational $\alpha$-relaxation times $\tau_t$ and $\tau_{r}$ are depicted in Fig. 3E,F. The rise of the $\alpha$ relaxation time $\tau_{r}$ for rotations sets in at lower $\phi$ than the rise in the corresponding translational $\tau_t$, and $\tau_{r}$ also exceeds the observation time at a lower $\phi$ than $\tau_{t}$. This reaffirms the presence of two different glass transitions. We also note that the correlators for translational relaxation at $\phi = 0.55$ and for rotational relaxation at $\phi = 0.46$ are well fitted, which confirms the fluid-like behavior at these $\phi$, which are the packing fractions of the fluid states closest to the transitions. 

A quantitative determination of the $\phi$ at which the glass transition for translation $\phi_c^t$ and rotation $\phi_c^{r}$ occur is obtained by the glass stability analysis of MCT \cite{Franosch1997} (cf. Materials and Methods). Performing such an analysis, we find that satisfactory fits to the data (Fig.~3E and F) can only be obtained assuming two different glass transitions $\phi_c^t=0.56$ and $\phi_c^{r}=0.50$. Therefore, also this analysis shows the existence of two separate glass transitions for rotation and translation in the experiment.

In summary, the analysis of the systems' dynamics reveals that we observed a density region where orientational motion in the sample was frozen while translational motion persisted. As this state lacks global nematic order, it is properly described as a liquid glass \cite{Letz2000}. 

\subsection*{Event driven Brownian dynamic simulations pointing to nematic correlations }%\protect\\}

\begin{figure}[htp]
\centering
\includegraphics[width=\columnwidth]{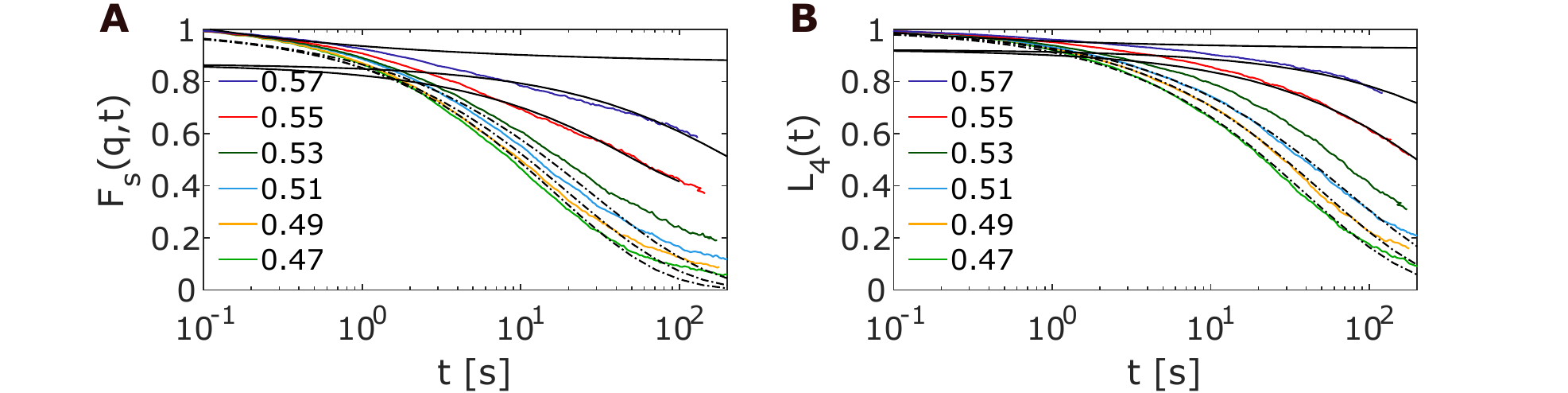}
\justify{\label{fig5} \textbf{Fig.~4. Glass transition analysis of simulation data.} (\textbf{A}) Translational $F_s(q,t)$ 
and (\textbf{B}) orientational $L_4(t)$ correlation functions from ED-BD simulations, where long-range 
nematic fluctuations are suppressed by rough walls. A MCT glass stability analysis (black lines for the two highest densities, cf.~Materials and Methods) 
finds a single glass transition density in this case.  The value  $\phi_c\approx 0.57$ is also supported
by a power law fit (with $\gamma=2.5$) to the Kohlrausch relaxation times for both translation and rotation.}
\label{fig:sims}
\end{figure}

The interpretation of experimental results obtained from colloidal suspensions often is complicated by the fact that the particles possess a certain polydispersity and might carry residual charges. Since these problems are non-existent in simulations, the latter are an important tool even if the number of particles which can be studied in this manner is typically rather small. To understand the origin of the formation of liquid glass in our system, we performed additional ED-BD simulations. We found that nematic order sets in quickly using monodisperse ellipsoids and periodic boundary conditions in accordance to previous simulation studies \cite{DeMichele2007,BautistaCarbajal2013}. Therefore, we tested different scenarios that could explain the formation of the experimentally observed glass states. First, since the experimental system is inherently polydisperse, we included polydispersity in our simulations. However, we found that a polydispersity of $5-15\%$, well above that of the experimental particle system used, did not affect the structure and dynamics of the system. Second, since Letz et al. \cite{Letz2000} predicted that the formation of a liquid glass is caused by long-wavelength fluctuations which may not be properly modeled in simulations due to finite size effects, we introduced rough walls in our simulations. The approach of confining glass-forming liquids has been used experimentally \cite{Naganamasa2011} and in simulations \cite{Mandal2014} for studies of polydisperse spheres. Indeed, we found that nematic order is suppressed in systems with rough walls. Results for the dynamics of this system are shown in Fig.~4 for $F_s(q,t)$ and $L_4 (t)$. Here, $L_4$ is shown as it exhibits glassy dynamics more clearly than $L_2$. The lines represent the MCT fitting curves obtained from Eqs.~\ref{eq:beta1} and \ref{eq:beta2}. Up to $\phi=0.53$ both correlation functions decayed to fluid states, while at $\phi=0.55$ and $\phi=0.57$ the system was already very close to the glass phase. However, in contrast to the experiments, the simulations still result in a single value for the glass transition of $\phi_c^t=\phi_c^{r} \approx 0.57$. 

While the discrepancy between the experimental system and simulations appears dissatisfying, it allows us to formulate a working hypothesis on the prerequisites for the formation of a glass and a liquid glass, respectively: It is known from experiments on suspensions of spherical colloids that the glass transition at the higher $\phi_c^t$ results from isotropic caging \cite{Stillinger2013}. This is accessible by simulation if nematic order is suppressed by the introduction of rough walls. Under this condition, the build-up of local neighbor shells leading to caging is possible already in small systems \cite{Chong2005}. The growing  fluid structure is reflected in the evolution of the structure factor $S(q)$ shown in Fig.~1D. By contrast, the formation of a liquid glass above $\phi_c^{r}$ and below $\phi_c^t$ requires long range correlations, which cannot be captured in the simulation box sizes currently accessible. This is especially true for shape anisotropic particles for which it has been shown that simulations are susceptible to finite size effects \cite{Dussi2018}. Solvent mediated interactions, which were neglected in the simulations, could also play a role, but have been shown to be of minor importance in the structural dynamics of spherical colloids \cite{Marenne2017}. We therefore conjecture that the discrepancy between experiment and simulations reveals that the nature of the observed liquid glass state is relying on long-range correlations.

\subsection*{Nematic precursor analysis}%\protect\\}

\begin{figure}[htp]
\centering
\includegraphics[width=\columnwidth]{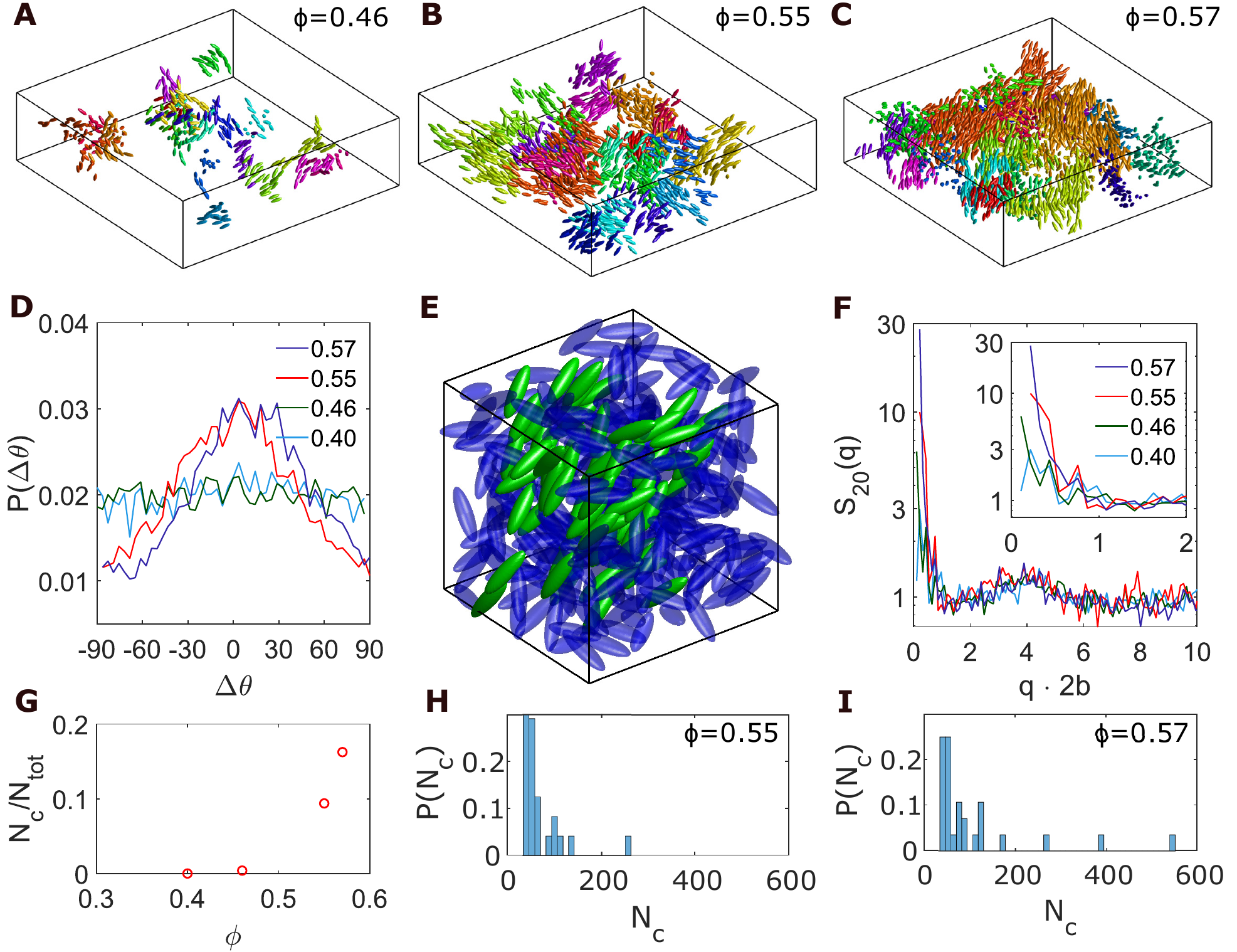}
\justify{\label{fig6} \textbf{Fig.~5. Analysis of nematic precursors.} (\textbf{A-C}) 3D representation of the 20 largest clusters with similar orientation, defined as described in the text, for different $\phi$ showing the existence of nematic precursors which are intercepting each other. Boxsize is $140\times140\times40\,\mu$m each. All particles are scaled by a factor of $0.8$ for better visibility.(\textbf{D}) Probability distribution $P(\theta)$ of the orientation of all ellipsoids in the system for different $\phi$. For increasing $\phi$, a favored orientation is emerging, indicating nematic precursors but no apparent nematic order. (\textbf{E}) The green particles belong to one cluster with similar orientation, while the surrounding blue particles are in an amorphous arrangement or belong to clusters of different orientations. The hindrance between the different precursors prevents global nematic order. Boxsize is $25\times25\times25\,\mu$m. (\textbf{F}) Orientational structure factor  $S_{20}(q)$ obtained from spatially correlating the orientations of all particles. Its dominating small-$q$ peak  records the growth of the nematic precursors. (\textbf{G})Normalized cluster size $N_c/N_{tot}$ for high $\phi$, where $N_c$ is the number of particles within a cluster and $N_{tot}$ the total number of particles within the image volume. The cluster size is increasing for higher $\phi$. (\textbf{H,I}) The distributions of cluster sizes for the two highest $\phi$.}
\end{figure}

To identify these long-range correlations and to test our hypothesis, we analyzed the spatial correlations in the samples. A close inspection of the imaging data revealed clusters of similarly oriented particles in the system. Two particles were assigned to the same precursor cluster, if they were next neighbors sharing a face of their individual Voronoi-cells and had an angular difference in orientation which was less than $\Delta\alpha \leq 20^{\circ}$. The choice of $\Delta\alpha$ was based on the half-width of the peak in the probability distribution $P(\Delta\theta)$ of the orientation of all ellipsoids in systems at high $\phi$  (Fig. 5D). Clusters were defined as nematic precursors, if they contained more than 30 particles. The choice of 30 particles was based on the largest cluster sizes found for $\phi=0.40$, which corresponds to an isotropic sample. Since the clusters do not form nematic regions, we term them nematic precursors. Fig. 5A-C depicts the 20 largest nematic precursors for the three highest $\phi$ studied. In general, we found that the number of particles in nematic precursors increased with growing $\phi$, as shown in Fig. 5G-I. Also the cluster size distributions became broader with increasing packing fraction. Our findings show that with $\phi$ increasing to densities where we find glassy states, more and more particles are found in local structures with a peculiar order. In these, no nematic order is detected even on small length scales, because other particles and clusters intersected the nematic precursors. This is exemplified in Fig. 5E, where the green particles belong to a nematic precursor whereas the blue particles are disordered surrounding particles intersecting the cluster. In the vicinity of the glass transition, particles tended to possess similar orientations (see Fig. 5D), but the nematic order parameter remained negligible (see Fig.~2). Nematic order formation is found only for particles with larger aspect ratios \cite{Roller2020}. The existence of the nematic precursors is also visible in the orientational structure factor, $S_{20}(q)$, which reflects the density fluctuations of quadrupolar symmetry, $\varrho_{l=2,m=0}({\bf q})$ proportional to the spherical harmonic for $l=2$ and $m=0$ \cite{Letz2000} (see Supplementary Materials). It is sensitive to the growth of nematic structures, which shows up as a peak for small wavevectors, and has been identified as correlation driving the glass transition at $\phi^r_c$  in MCT \cite{Letz2000}. Fig. 5F (note the log-linear plotting) shows that the large-$q$ peak in $S_{20}(q)$, which records the orientational alignment of neighboring particles, remains rather unchanged while, upon increasing $\phi$, the nematic precursors cause a growing peak at small wavevectors. Its width measures the average size of the clusters, which exceeds the average particle separation by roughly a decade.

Our finding of a liquid glass state and of nematic precursors in a system of colloidal ellipsoids in 3D is in stark contrast 
to experimental and theoretical results for colloidal ellipsoids confined in 2D \cite{Mishra2013, Zheng2014}. While for an 
aspect ratio of $a/b=3.5$ the 2D studies also show the appearance of a state in which rotational dynamics is frozen and 
translational mobility persists, the structural order of this state is very different from the 3D case, as in 2D domains with 
high nematic order are found. It has been suggested that the driving force for these nematic domains in the 2D system could be 
long-wavelength fluctuations, as first described by Mermin and Wagner \cite{Zhang2019}. However, as Mermin-Wagner fluctuations 
are absent in 3D, we conclude from our experiments and simulations that in the 3D case long-wavelength fluctuations are 
supported by long-range nematic correlations, induced by nematic precursors that are intersected by differently oriented 
particles. Thus, nematic precursors appear to be prerequisite for the formation of bulk liquid glass states.

%%%%%%%%%%%%%%%%%%%%%%%% END OF RESULTS %%%%%%%%%%%%%%%%%%%%%%%%%%%%%

\section*{Conclusions}

Hard ellipsoidal particles are a relatively simple system posing an entropic packing problem only slightly more intricate than the one of spheres. Considering this, the richness of the states diagram of equilibrium and kinetically arrested states is striking. As is well known since Onsager's work on thin needles, above a specific density hard elongated particles align locally for entropic reasons. Unexpectedly, for an intermediate aspect ratio of the ellipsoidal colloids,  local alignment gets frustrated on intermediate length scales and ramified, differently oriented regions result. Our simulations show that polydispersity plays no major role and we thus expect the formation of glass driven by differently aligned cooperative regions to exist in other glass-forming systems as well. This effect thus needs to be considered in the active concerted efforts to form structured materials from colloidal constituents. Our findings suggest that obtaining high order will require strong enthalpic contributions. The emerging nematic precursors contain hundreds of particles and it is the mutual obstruction of these cooperative regions that leads to the formation of liquid glass. We showed that the critical long-range fluctuations connected with nematic ordering, which is a weakly first order phase transition, are involved in the formation of this novel glassy state. This should be contrasted with the familiar glass transition which tracks the crystallization line, where a strongly first order phase transition takes place. Thus, the liquid glass state may give the long sought paradigm where the interplay between equilibrium critical correlations and critical slowing down versus glass-formation can be studied microscopically. This promises to shed light on the origins of dynamic heterogeneity in molecular systems. Additionally, it could also result in an intriguing venue for future studies concerning the formation and evolution of topological defects. While on a molecular scale, these dominate the phase-ordering dynamics in liquid crystals \cite{Bradac2011}, they also serve as a model system for the evolution of matter in the universe according to the Kibble-Zurek mechanism \cite{Chuang1991}. In both cases, the glassy arrest observed in our results will have an important influence on the resulting structures.

%%%%%%%%%%%%%%%%%%%%%%%% END OF DISCUSSION  %%%%%%%%%%%%%%%%%%%%%%%%%%%%%

\section*{Materials and Methods}
\subsection*{Experimental methods}
In a first step, spherical PMMA/PMMA core-shell particles dyed with different fluorophores in the cores and the shells were synthesized according to a route described in \cite{Antl1986, Roller2018}. As fluorophores 
for the cores we used the Bodipy dye 
((4,4-difluoro-8-(4-methacrylatephenyl)-3,5-bis-(4-methoxyphenyl)-4-bora-3a,4a-diaza-s-indacene),
which was synthesized according to \cite{Baruah2005}. The dye Quasar 670 (LGC Biosearch Technologies) 
served as a marker for the shells and was used in form of a 4-vinylbenzylester. 
For core particle synthesis, the dye was linked to methacrylic acid (Fluka) and copolymerized. 
The cores were cross-linked using ethylene glycol dimethylacrylate. The resulting particle 
cores had a diameter of 1$\;\mu$m. Several seeded dispersion polymerization steps were then used to grow a PMMA shell 
of the desired thickness onto these cores. The PMMA in the shell material was not cross-linked. 

In a second step, the particles were thermo-mechanically stretched as reported by Keville et al. 
\cite{Keville1991} with modifications for the PMMA/PMMA core-shell particles as described in 
\cite{Klein2014, Klein2013, Klein2015}. After the stretching, the ellipsoidal particles were 
restabilized with polyhydroxystearic acid (PHSA) and dispersed  
in a mixture of 85:15 (w/w) cyclohexylbromide and cis-decalin to match both density and refractive 
index of the PMMA colloids \cite{Royall2012}. Since PMMA particles are known to charge in organic solvents, tetrabutylamoniumbromide was added to screen the charges, 

The resulting suspensions were filled into a home-built sample chamber. Samples were studied 
using a confocal laser scanning microscope (TCSP5, Leica Microsystems) equipped with a resonant scanner (8 kHz, bidirectional 
mode) and a glycerol objective (63x, NA 1.30). An argon laser (wavelength $\lambda=514.5\,$nm) and a helium neon laser 
($\lambda=632,8\,$nm) were used for excitation of the dyes. The temperature of the whole microscope system was stabilized at 
$23.0\pm0.1^{\circ}$C (Ludin Cube 2, Life Imaging Services), since at this temperature the density match between particles and 
solvent was optimal. Volume fractions up to $\phi=0.46$ were prepared by centrifugation at higher temperature and removing 
solvent. The highest $\phi$ were obtained by slightly raising the temperature in a dense sample directly in the 
measurement chamber for a few days to achieve density mismatch. This led to the slow enrichment of particles at the bottom of 
the sample.

The temperature was then changed back to $23^{\circ}$C. After placing a sample onto the microscope it was allowed to 
equilibrate for $15$h to minimize drift effects. Remaining drift, which in all cases was small compared to the particle 
movement, was corrected with a routine applied after detection and tracking. 
To minimize wall effects, all measurements were performed in a depth of $30-100 \mu$m into the sample. Positions and 
orientations of all particles in one image volume (144$\,\mu$m$\times$144$\,\mu$m$\times$40$\,\mu$m for structural analysis 
and 144$\,\mu$m$\times$36$\,\mu$m$\times$20$\,\mu$m for measurements on the dynamics) containing more than 6000 particles were 
identified and linked to trajectories using self-written and tested Matlab algorithms. The core-shell structure facilitates 
tracking of the ellipsoid position and orientation with accuracies of $60\,$nm and $5^\circ$, respectively \cite{Roller2018} 
(Fig. 1B). Experimental time-scales were limited by the bleaching of the fluorophores after long laser exposure. 

Volume fractions are the pivotal variable influencing the structural dynamics of the colloidal suspensions. For their 
determination, particles were detected and counted in each measured sample and the related Voronoi-volume was calculated to 
obtain the exact number fraction $\rho$. Multiplying this with the volume of a particle gives $\phi$:
\begin{equation}
\phi = \frac{4}{3}\pi \left(\frac{d}{2}\right)^3 \rho.
\end{equation}
The determination of particle volumes, however, is intricate since PMMA particles are known to swell in CHB. Therefore, we determined the particle diameter by calculating the pair correlation function for a dense 
sample of spherical particles before stretching. The volume of a particle is conserved upon stretching, which was proven by 
crosschecking the obtained value with the width of an ellipsoid acquired by the pair correlation of the elliptical particles.

\subsection*{Glass stability analysis}

The values for the glass transitions for translation $\phi_c^t$ and rotation $\phi_c^{r}$ are found by the glass
stability analysis of MCT in which the collection of all measured correlation functions is used. Close to the glass transition  $\phi_c$, a critical decay onto the plateau occurs 
in a correlation function $\Phi_j(t)$ (e.g. the translation or the rotation correlator functions $F_s(q,t)$ and $L_n(t)$, 
respectively). Using first order correction terms \cite{Franosch1997}, the decay is given by
\begin{equation}
\Phi_j^c(t) = f_j^c + h_j\left(\left(\frac{t}{t_0}\right)^{-a} + \left(k_j+\kappa(a)\right)
\left(\frac{t}{t_0}\right)^{-2a}\right) \label{eq:beta1}
\end{equation}
where $t_0$ and $\kappa(a)$ are system dependent constants and $h_j$, $k_j$ and $f_j^c$ are parameters, which depend on the 
correlator $j$. 

\begin{figure}[htp]
\centering
\includegraphics[width=\columnwidth]{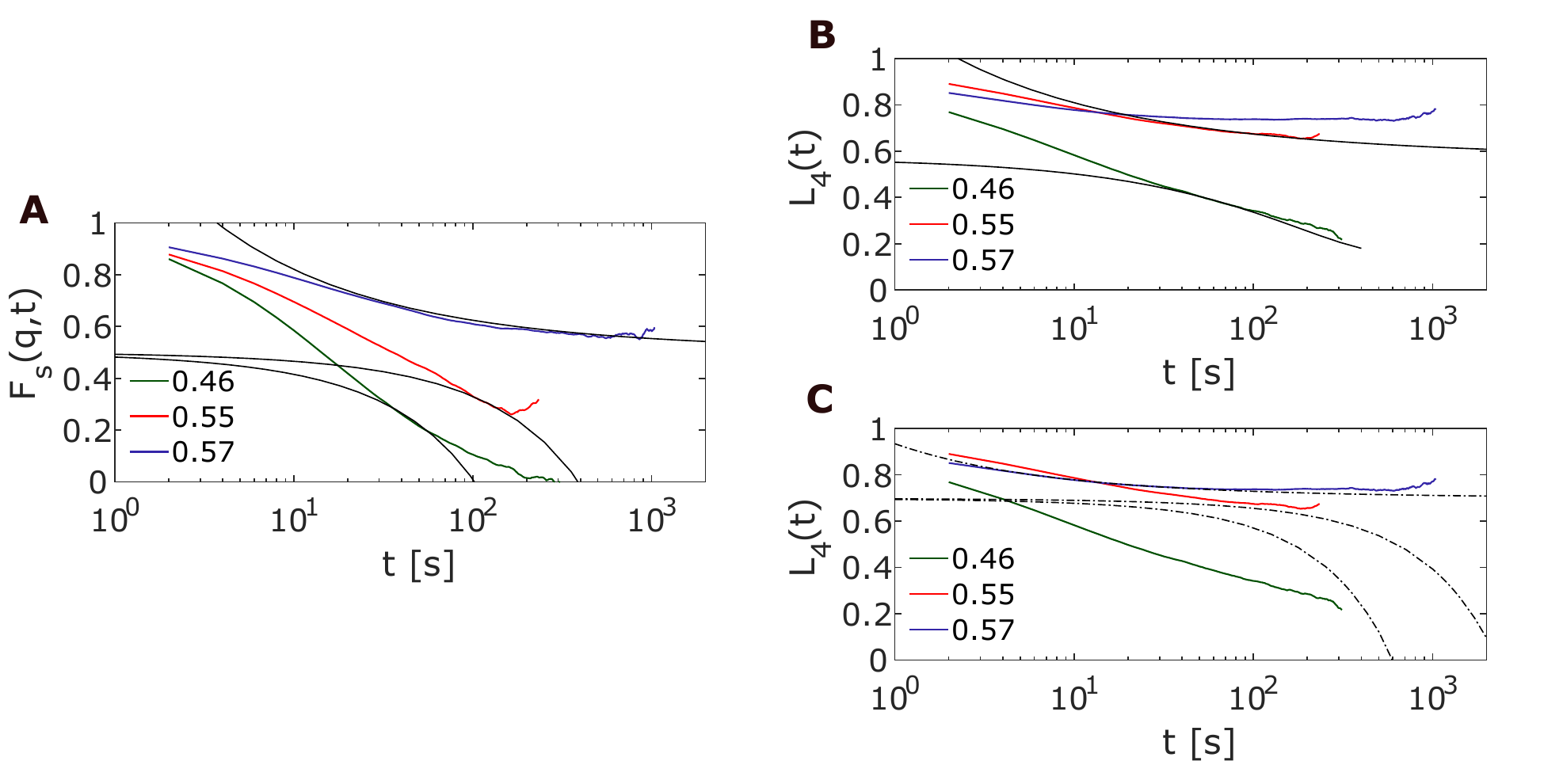}
\justify{\label{fig4}\textbf{Fig. 6. Glass stability 
analysis of translational (\textbf{A}) and (\textbf{B}) rotational relaxations:} Data from experiment (colored lines) 
and fits (solid lines) according to equations \ref{eq:beta1} and \ref{eq:beta2}. Two 
different glass transitions for translation  $\phi_c^t=0.56$ and rotation $\phi_c^{r}=0.50$ were 
necessary to fit the data. In panel (\textbf{C}), the grey dash-dotted lines indicate the result, 
when an identical transition density was assumed for translation and rotation. For this condition, poor agreement 
with the data at $\phi=0.46$ was observed.}
\end{figure}

Below $\phi_c$, the late plateau and the decay from the plateau is given by the von Schweidler law:
\begin{equation}
\Phi_j(t) = f^c_j - h_j\left(\left(\frac{t}{\tau_{\phi}}\right)^{b} + k_j\left(\frac{t}{\tau_{\phi}}\right)^{2b}\right) 
\label{eq:beta2}
\end{equation}
where $\tau_{\phi}$ is a time scaling depending sensitively on $\phi$
\begin{equation}
\tau_{\phi}= \frac{t_0}{B^{1/b}}\left(C\frac{(\phi_c - \phi)}{\phi_c}\right)^{-\gamma}
\end{equation}
and determines the final relaxation time $\tau$ close to 
and below $\phi_c$. Importantly, the exponents $a$, $b$ and $\gamma$ are related and determine all 
universal aspects close to the glass transition. Here, we employ the exponents $a=0.3$, $b=0.6$ (therefore $\gamma=2.5$) and 
$B=0.84$ and $C=1.54$ to represent a repulsive glass-forming system\cite{Franosch1997}. For the glass stability analysis, both 
translation and rotation correlator functions were fitted simultaneously using Eqs.~(\ref{eq:beta1}) and  (\ref{eq:beta2}) for 
the three highest volume fractions $\phi=0.46$, $\phi=0.55$, and $\phi=0.57$ considering all measured $F_s(q,t)$ in the range 
$5.18\le q \cdot 2b\le7.90$ and $L_n$ for $n=2,4$. Results are shown in Fig. 6A,B (again $L_4$ is shown because 
the glassy two-step relaxation is more visible than in $L_2$). The best fitting results were obtained 
by keeping $\kappa(a)=1$ and $t_0 = 0.8$ fixed and having $f_j$, $h_j$, $k_j$, and  $\tau_{\phi}$ as free fitting parameters. 
Satisfactory fits to the data were only obtained assuming two different glass transitions $\phi_c^t=0.56$ and 
$\phi_c^{r}=0.50$. The grey dotted lines in Fig. 6C indicate the result when only one glass transition density was 
assumed for translation and rotation. Additional background on the MCT states diagram and the stability analysis is given in the Supplementary text.

\subsection*{Simulation details}

We performed event-driven Brownian dynamics (ED-BD) simulations \cite{Scala2007} 
that model overdamped Brownian dynamics of hard ellipsoids. The system consisted of $N$ non-overlapping 
ellipsoids in a simulation box of length $L$ and volume $V = L^3$ that was varied depending on the 
desired $\phi$. Similar to De Michele et al. \cite{DeMichele2007}, we set the moment of 
inertia  to $I=2mb^2/5$. For volume fractions of $\phi = 0.05, 0.1, 0.2$, ellipsoids were placed 
randomly in the simulation box, while for volume fractions between $\phi = 0.4$ and $\phi=0.57$, the 
Lubachevsky-Stillinger \cite{Lubachevsky1990} technique was used. All systems were equilibrated by 
Newtonian dynamics allowing translational and rotational correlation functions to decay to zero. Two 
types of boundary conditions were modeled. The first case corresponds to systems with periodic boundary 
conditions (PBC) with $N=504$. The second case corresponds to 
systems confined within rough walls. For the confined systems, we 
started from a random configuration of $1231$ ellipsoids and pinned the outermost particles in the 
range $0 < d < \Delta$ and $L - \Delta < d < L$ to serve as the wall particles. The remaining $500$ 
particles were free to move.  The ED-BD model was first benchmarked for a dilute system of $\phi=0.05$, 
where we reproduced the ratio of translational and rotational diffusion coefficients found in the 
experiments: $D_{\mbox{\tiny{rot}}} b^2/ D_{\mbox{\tiny{trans}}}=1.25$ setting the ratio of Brownian 
time steps  $\Delta t_{\mbox{\tiny{rot}}}/\Delta t_{\mbox{\tiny{trans}}} = 0.331$ 
(with $\Delta t_{\mbox{\tiny{rot}}}= 2I/k_BTD_{\mbox{\tiny{rot}}} = 0.0265$ short enough to give 
angular diffusion).  This ratio was then applied to all $\phi$. At all densities, the time 
scale when matching to the experiments is taken from $D_{\mbox{\tiny{trans}}}(\phi)$. 
All presented results 
were averaged from 20 initial configurations of the ellipsoids.

%%%%%%%%%%%%%%%% END OF MATERIALS AND METHODS %%%%%%%%%%%%

% Your references go at the end of the main text, and before the
% figures.  For this document we've used BibTeX, the .bib file
% scibib.bib, and the .bst file Science.bst.  The package scicite.sty
% was included to format the reference numbers according to *Science*
% style.

%BibTeX users: After compilation, comment out the following two lines and paste in
% the generated .bbl file. 

\bibliography{scibib}

\bibliographystyle{Science2}

\section*{Acknowledgments}

We gratefully acknowledge the financial support by Deutsche Forschungsgemeinschaft, SFB1214, TP B5. We thank S. Schütter and 
F. Rabold for particle synthesis, Dr. P. Pfleiderer for his contribution on particle detection, Dr. M. Gruber for calculations 
on the schematic MCT model and Profs. Dr. H. Lekkerkerker and Dr. G. Maret for helpful discussions. J.-M. M. acknowledges funding provided by the Alexander von Humboldt Foundation.

\newpage

\section*{Supplementary materials}
Supplementary text to the following figures:\\
Fig.~S1: Superposition of correlation functions  $F_s(q,t)$ and $L_n(t)$ from experiment\\% plotted against $\log(\Delta t n^2)$ \\
Fig.~S2: Mean squared displacement, mean squared angular displacement and diffusion coefficients from experiment for different $\phi$ \\
Fig.~S3: Correlation functions $F_s(q,t)$ and $L_n(t)$ and nematic order parameter $Q$ of the Brownian simulated system with periodic boundary conditions \\
Fig.~S4: Correlation functions $F_s(q,t)$ and $L_n(t)$ of the Molecular dynamics simulated system with walls \\
Fig.~S5: $S_{20}(q)$ of the Brownian simulated system\\
Fig.~S6: State diagram of a schematic MCT model\\

%\newpage
\setcounter{equation}{0}
\renewcommand{\theequation}{S\arabic{equation}}

\subsection*{Control experiments and fit parameters:}

As expected for dilute systems,   the curves for all determined correlation functions can be superimposed for volume fractions $\phi$ below the glass 
transition. This requires that the density correlators 
$F_s(q,t)$ for several values of $q$ are plotted versus $\log(\Delta t q^2)$ and the orientational correlators 
$L_n(t)$ are plotted against $\log(\Delta t n^2)$) (Fig. S1).

\begin{figure*}[h!]
\centering
\includegraphics[width=\columnwidth]{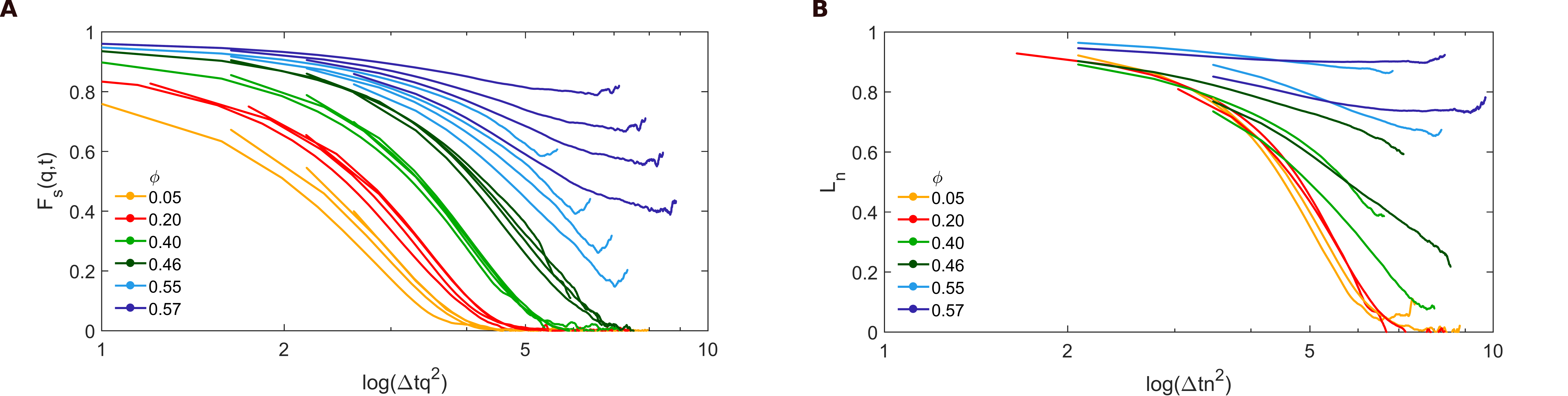}
\justify{
\textbf{Fig.~S1. Superposition of the correlation functions.} For packing densities below the glass 
transition, the values obtained for the density correlator $F_s(q,t)$ at different values of $q$ can be superimposed 
if they are plotted versus $\log(\Delta t q^2)$ (\textbf{A}). The four depicted wavevectors lie in the range $5.18< q\cdot2b<7.90$. Superposition is also possible when the orientational correlators $L_n(t)$ 
are plotted against $\log(\Delta t n^2)$ (\textbf{B}) for $\phi<\phi_c$. Here, correlators for $n=2$ and $4$ are shown.}
\end{figure*}

The data on the systems's dynamics shown in the 
main text were measured several times for each volume fraction. 
A combination of all the measured data is shown in Fig. S2A,B, where we plotted the mean squared 
displacement (MSD) $\langle\Delta r^2\rangle$ and the mean squared angular displacement (MSAD) $\langle\Delta 
\theta^2\rangle$ for all  measurements. Since the data nicely collapse onto one curve for each $\phi$, for clarity plots containing just single curves are shown in the main text. To evaluate the homogeneity of 
the particle movements along the axes parallel $\langle\Delta r^2\rangle_{||}$ and perpendicular $\langle\Delta 
r^2\rangle_{\perp}$ to the orientation axis of an ellipsoid, we separately analyzed the respective MSDs (Fig. S2C,D). From fits 
to the data, we obtained short time diffusion coefficients $D_{||}$ and $D_{\perp}$. As expected, particles tend 
to move faster along the orientation axis than perpendicular to it as is shown in the plot $D_{||}/D_{\perp}$. While 
this ratio slightly increases with $\phi$ and have a maximum in the liquid glass state at $\phi=0.55$, they 
hardly deviate from the values expected for a freely diffusing ellipsoidal particle indicated by the dashed line in Fig. S2E \textit{(39)}. 

\begin{figure*}[h!]
\centering
\includegraphics[width=\columnwidth]{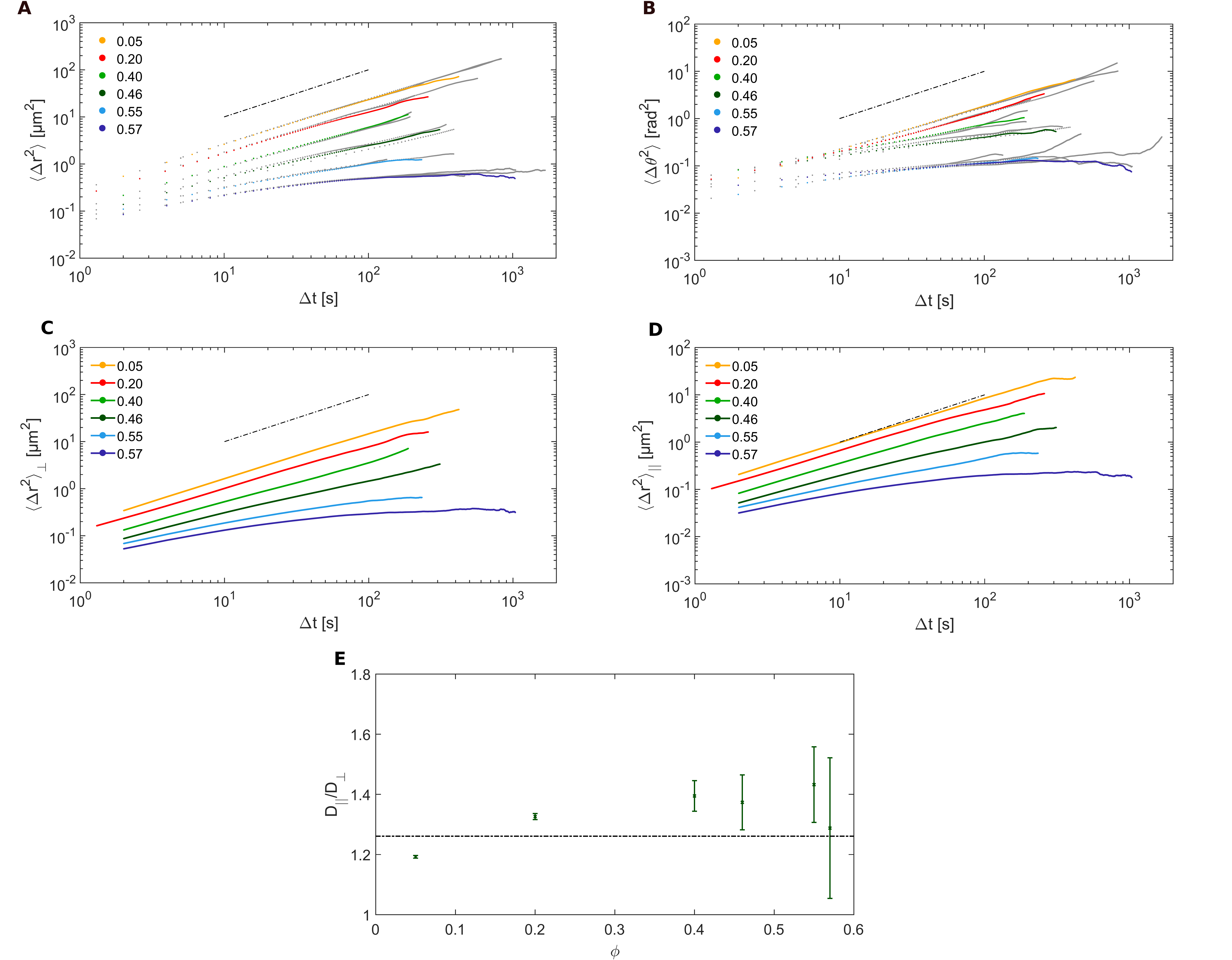}
\justify{
\textbf{Fig.~S2. Experimental data showing the mean squared displacement (MSD) $\mathbf{\langle\Delta r^2\rangle}$ (\textbf{A}) and mean 
squared angular displacement (MSAD) $\mathbf{\langle\Delta \theta^2\rangle}$ (\textbf{B}) for all measured samples (grey lines).} Very good coincidence is found 
for subsequent measurements at the same $\phi$. Therefore only one measurement for each $\phi$ is shown in the main text which appear to be good representatives (colored lines). The dotted line indicates the slope $m=1$ which shows the free diffusive behaviour.  Additional calculation of the mean 
squared displacement perpendicular $\langle\Delta r^2\rangle_{\perp}$ (\textbf{C}) and parallel $\langle\Delta r^2\rangle_{||}$ (\textbf{D}) to 
the orientation axis of an ellipsoid from the experiments. (\textbf{E}) The ratio of the obtained diffusion coefficients, the vertical dashed line illustrates the value for a free particle derived in Ref.~\textit{(39)}.  }
\end{figure*}

As described in the main text, we fitted the obtained correlation functions to the Kohlrausch-function (Eq.~(\ref{eq:KWW}). Only curves which showed a clear 
decay within the measured time window were fitted; representative examples are shown in Fig.~3C,D. Fit parameters are collected in Table 1. \\

\begin{table}
\begin{tabular}{l|lllll}
$\phi$ & 0.05 & 0.2 & 0.4 & 0.46 & 0.55 \\
\hline
translation ($q = 3.2 \mu m^{-1})$ & 0.95774 & 0.88347 & 0.81916 & 0.62496 & 0.39930 \\
rotation ($n=4$) & 0.77396 & 0.74768 & 0.53068 & 0.33530
\end{tabular}
\textbf{\caption{\normalfont{Fit parameters for the fits shown in Fig. 3 B and D.}}}
\end{table}

\subsection*{Details on the simulation results:}

Fig.~S3 shows the correlation functions, $F_s(q,t)$, $L_2(t)$, $L_4(t)$ and the order parameter $S$ for the systems with periodic boundary conditions. Similar to De Michele et al. \cite{DeMichele2007}, the isotropic-nematic threshold was set to $S = 0.3$, hence the nematic
transition is at $\phi \approx 0.49$. For nematic systems, $F_s(q, t)$ monotonically decreases to 0
without significant stretching, while $L_2(t)$ tends to form a plateau that corresponds to the orientational ordering. We verified the  formation of a nematic state in these systems by calculating the plateau
height of $L_2(t)$ and the final value of $S^2$ \cite{Eppenga1984},
\label{eq:L2Q2}
\begin{equation}\label{eq:s2}
\lim_{t\to \infty}L_2(t) = S^2
\end{equation}

For the nematic systems obtained in the simulations, the final values of $L_2(t)$ are close to the $S^2$ values despite the finite time and finite size of the box. %By contrast, in the liquid glass state, plateau values of $L_2(t)$ much higher than the corresponding $S^2$ values of $\approx 0.2$ were observed.

Fig.~S4 shows $F_s(q, t)$, $L_2(t)$ and $L_4(t)$ for the systems with rough walls. In this figure, the
characteristic plateau formation for glasses can be observed. Unlike the systems with
periodic boundary conditions (Fig.~S3), the corresponding $S$ values for the systems with rough
walls are all below 0.3. Plateaus are clearly observed starting at $\phi= 0.55$. The 
$L_2(t)$ plateau value is $\approx 0.8$ and Eq.~(\ref{eq:s2}) is no longer obeyed. Instead, the plateau values
are used for the von Schweidler fits for glassy systems as discussed in the main text.

\begin{figure}[h!]
\centering
\includegraphics[width=\columnwidth]{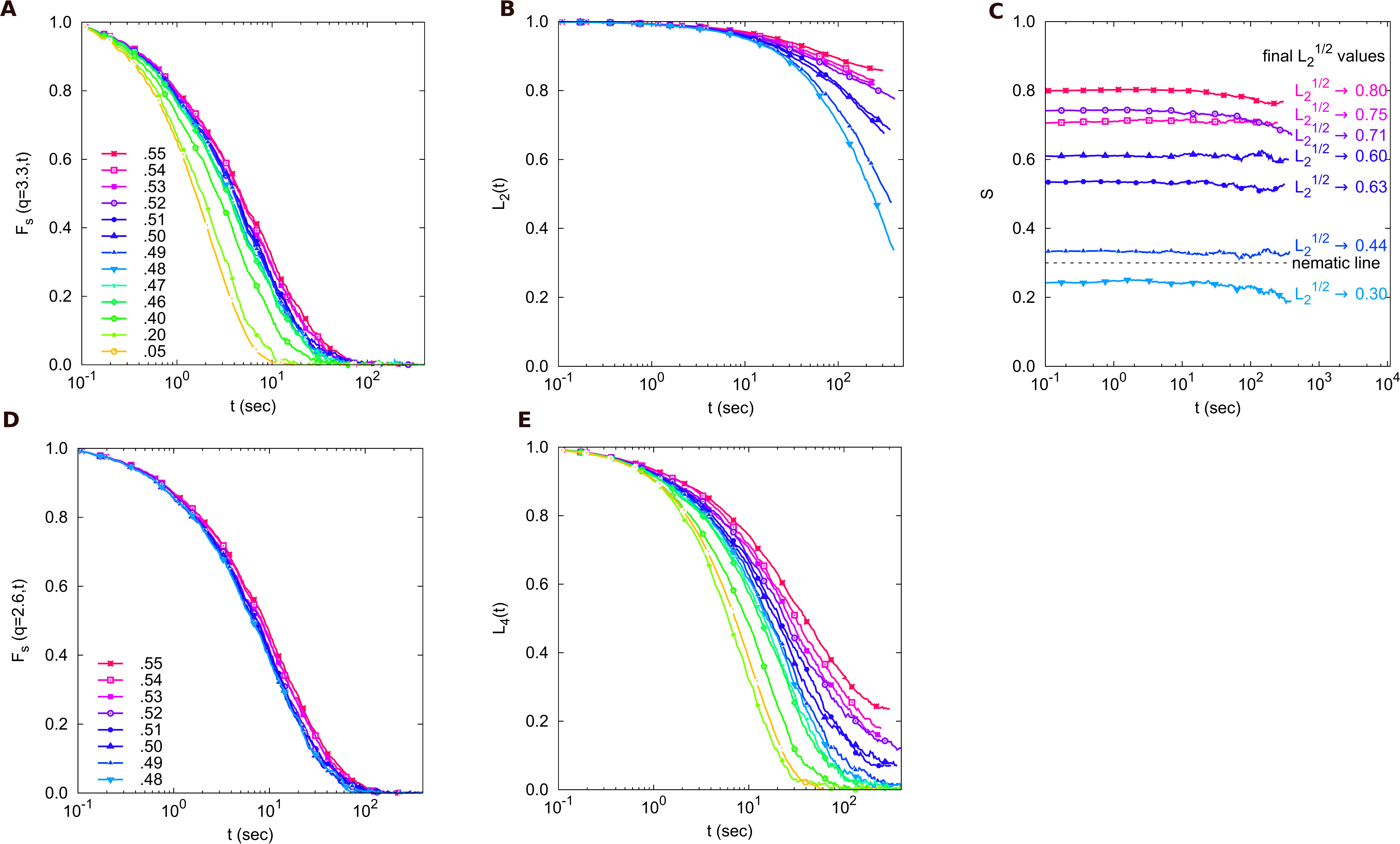}\\%
\justify{
\textbf{Fig. S3. Data from BD simulations with periodic boundary conditions for $\mathbf{F_s(q,t)}$ (\textbf{A}), $\mathbf{L_2(t)}$ (\textbf{B}) and $\mathbf{S}$ (\textbf{C}) where the plateau formation is attributed to nematic ordering starting at $\mathbf{\phi=0.49}$.} Figure legend of part \textbf{A} also applies to \textbf{B,C}. Additional data for $F_s(q,t)$ (\textbf{D}) and $L_4(t)$ (\textbf{E}) }
\label{fig:bd-pbc}
\end{figure}

\begin{figure}[h!]
\centering
\includegraphics[width=\columnwidth]{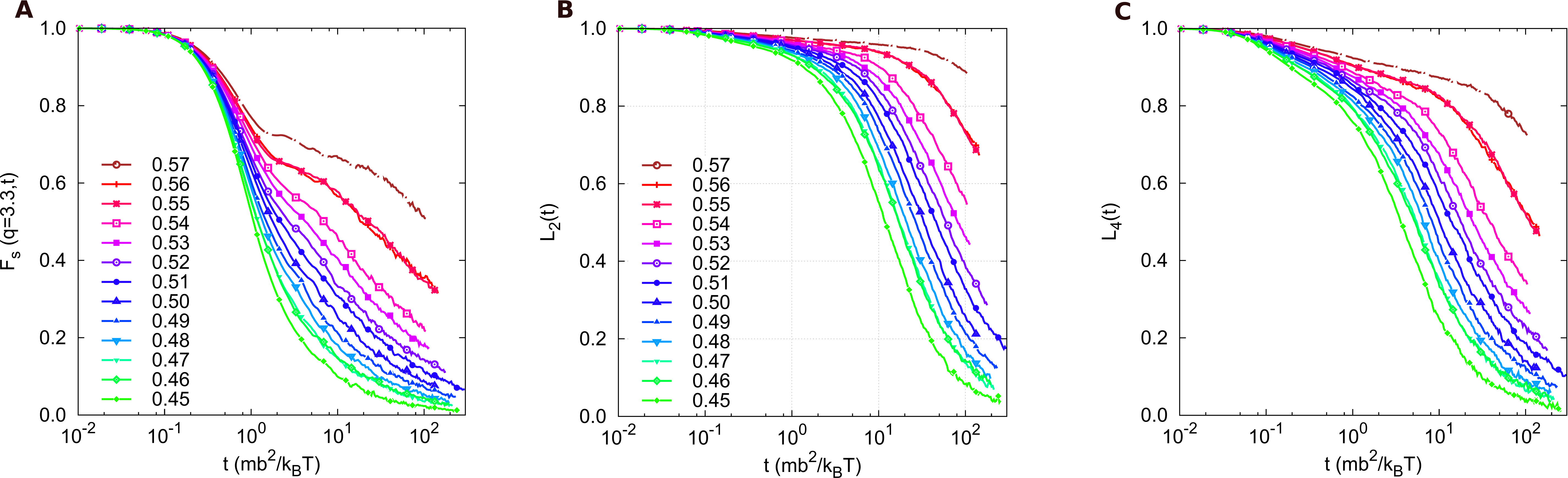}\\%
\justify{
\textbf{Fig. S4. Data from MD simulations with rough walls for $\mathbf{F_s(q,t)}$ (\textbf{A}), $\mathbf{L_2(t)}$ (\textbf{B}) and $\mathbf{L_4(t)}$ (\textbf{C}) where the plateau formation for high packing fractions is attributed 
to glass formation at $\mathbf{\phi \approx 0.56}$.} Figure legend of part \textbf{B} also applies to \textbf{C}.}
\label{fig:md-walls}
\end{figure}

$S_{lm}(q)$ is an orientation-dependent structure factor that can be used to analyze the correlation of the orientations of the ellipsoids. It is defined as 
\begin{equation}
\label{eq:S20}
S_{lm} (q) = \frac{1}{N} \langle \rho_{lm}^*(q) \rho_{lm}(q) \rangle,
\end{equation}
with the microscopic density defined as 
\begin{equation}
\rho_{lm}(q) = \sqrt{4 \pi} i^l \sum_{j=1}^N e^{i{\bf q} \cdot {\bf r}_j } Y_{lm} ({\bf \Omega}_j),
\end{equation}
where $Y_{lm} ({\bf \Omega})$ is the spherical harmonic function for Euler angles ${\bf \Omega}(\theta,\phi)$. Note that we only consider the diagonal terms of the orientation-dependent structure factor. The results for \mbox{$l=2, m=0$} in the simulations (using periodic boundary conditions) are shown in Fig. S5. The transition to a nematic phase in the simulations is reflected in $S_{20}$ in two ways. First, for small $q$-vectors, $S_{20}$ increases as $\phi$ increases, signaling the formation of long-range order that is limited to the size of the simulation box. Second, the neighbor peak of $S_{20}$ starts to become more visible and shifts to the right as $\phi$ increases. That is, the favored alignment of neighboring ellipsoids is such that their axes of symmetry ($a$) are parallel to each other.

\begin{figure*}[h!]
\begin{center}
\includegraphics[width=0.3\columnwidth]{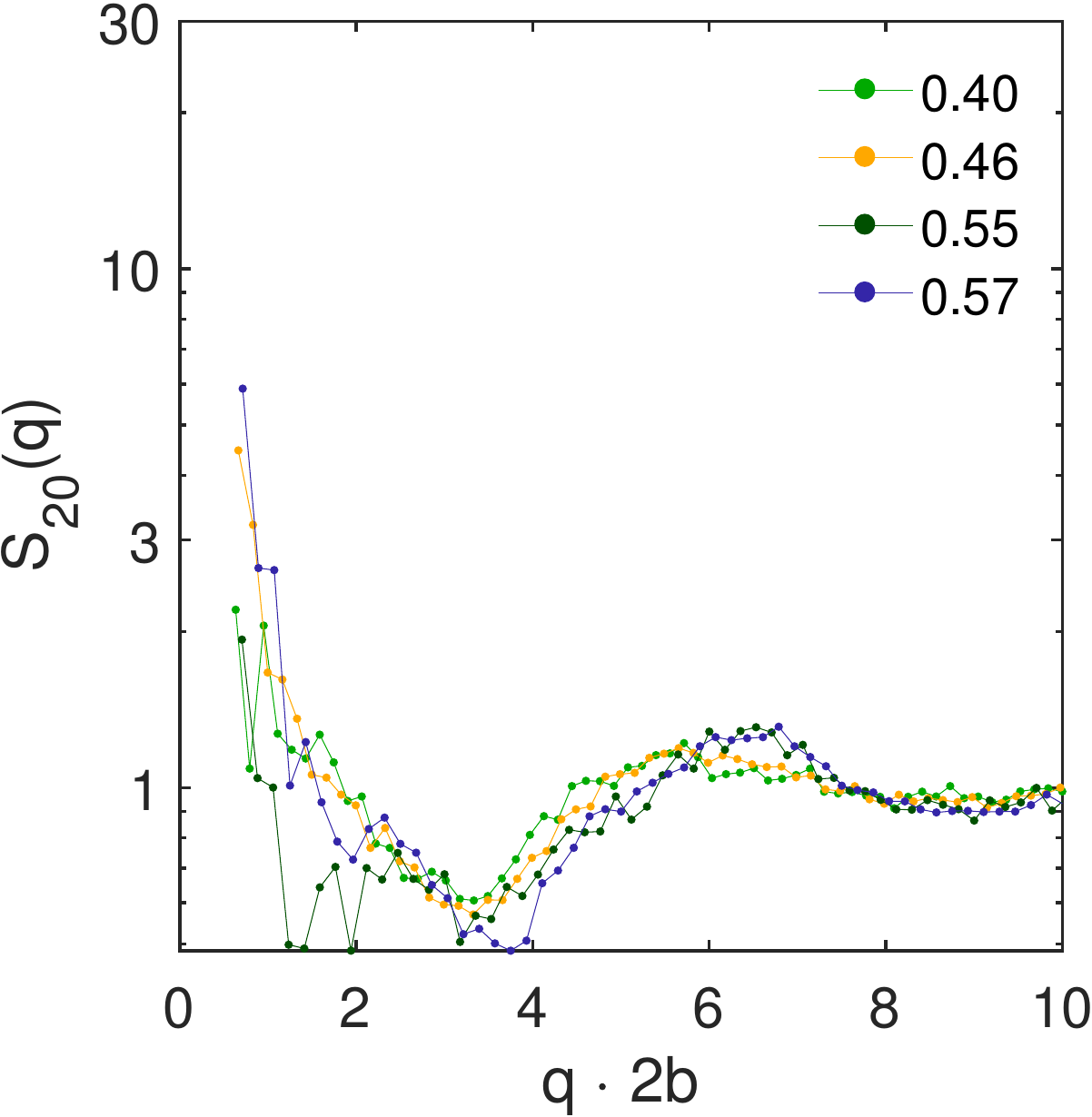}\end{center}
\justify{
\textbf{Fig. S5. Simulation counterpart of Fig. 5F of the main text.} Calculated $\mathbf{S_{20}(q)}$ shows the nematic transition for $\mathbf{\phi>0.46}$, indicated by the right shift of the first peak.}
\end{figure*}

\subsection*{Mode coupling theory}
We argue that the region of liquid glass in phase space is bounded by two glass transitions at fixed 
aspect ratio, providing the basis for our MCT glass stability analysis in the main text and extending the previous MCT calculations \cite{Letz2000}; there the possibility of a glass-to-glass transition was not explored.

 The transition of a fluid 
of hard ellipsoids to a liquid glass was found by MCT  in a fully microscopic 
calculation. It was shown to be driven by long-range nematic correlations which arise close to the 
equilibrium isotropic-nematic transition (see Fig.~5F recording these correlations in the samples). 
The transition line meets another line of fluid to glass 
transitions, which extends to the MCT hard sphere transition for aspect ratio approaching unity. 
It is driven by an increase of liquid short range structure seen in $S(q)$ ('cage effect'; see Fig.~1D). As MCT glass transitions 
are fold bifurcations in a nonlinear algebraic system, transition lines do not stop when they 
meet but rather intersect generically. A schematic model shows that the 
latter transition line extends into the glassy region above the transition to the liquid glass. 
There it gives a line of liquid glass to glass transitions which is in agreement with the experimental 
observations. 

\begin{figure}[htp]
\centering
\includegraphics[width=\columnwidth]{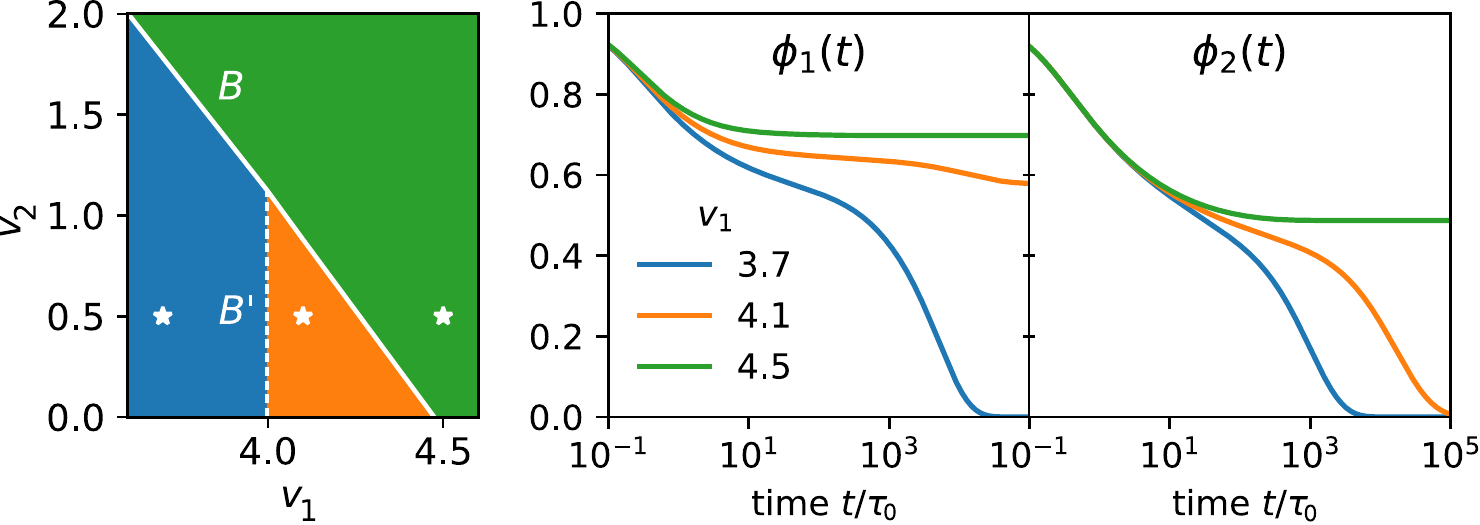}
\justify{
\textbf{Fig.~S6 Schematic MCT model capturing the generic glass transition scenario of two modes:} The left panel shows the states diagram at 
fixed $v_3=0.7$ and $w=3$, where two discontinuous bifurcations  exist.  They separate fluid ($f_1=0$, $f_2=0$, blue) from liquid glass 
($f_1>0$, $f_2=0$, orange) and glass ($f_1>0$, $f_2>0$, green) states. Parameters $v_1$ and $v_2$  mimick aspect ratio and density.  The other panels show the correlators: The middle panels shows 
$\Phi_1(t)$ modeling orientational motion (viz.~$L_n(t)$), and the right panel $\Phi_2(t)$ modeling translational motion (viz.~$F_s(q,t)$); the overdamped MCT equations 
of motion are solved  for the points marked by stars in the states diagram (precise values are $v_1=3.7$, 4.1, 4.5  at $v_2=0.5$).}
\end{figure}

Starting point is the model by Bosse and 
Krieger \textit{(40)} (BK). It describes the generic coupling of two degrees of freedom in the case of a single (discontinuous or 
generic \textit{(41)}) glass transition. Their correlators $\Phi_1(t)$ 
and $\Phi_2(t)$ shall correspond to $L_n(t)$ and $F_s(q,t)$, respectively. The $\Phi_i(t)$ obey Zwanzig-Mori equations with memory kernels $m_i(t)$ (for $i=1,2$) given as a general 
quadratic form.  The slowing-down of the correlators' relaxation arises from the feedback in the retarded friction kernels 
modeled by  $m_1(t)=v_1\Phi_1(t)^2+v_2\Phi_2(t)^2$  and $m_2^{\rm BK}(t)=v_3\Phi_1(t)\Phi_2(t)$. We generalize the model by including 
a parameter $w$ modeling the coupling of the second dynamical mode to itself; viz.~$m_2(t)=v_3\Phi_1(t)\Phi_2(t)+w\Phi_2(t)^2$.  This 
allows for a second generic glass transition. The two glass transitions of the model will correspond to the $B$ and $B'$ transitions  introduced by Letz et al. \cite{Letz2000}. The parameters $v_1$ and $v_2$ encode  the increasing orientational friction arising due to slow orientational and translational motion, respectively. Thus $v_1$ should correlate with the  aspect ratio and $v_2$ with the density. The cross-coupling term $v_3$ parametrizes the translational friction arising from rotation-translation coupling, while $w$ captures the feedback within the translational motion only. The glass parameters 
$f_i=\Phi_i(t\to\infty)$ obey the equations $f_i/(1-f_i)=m_i(t\to\infty)$, where glass transitions appear as 
bifurcations. Since the model lacks a quadratic coupling of the first mode into the second kernel, it contains a 
type $B'$ transition at $v_1=4$ and $v_2$ small enough, where $f_1$ jumps from zero to a finite value, while $\Phi_2(t)$ remains fluid 
like. For parameter sets with a second transition from fluid to glass, which is continuous in the BK-model ($w=0$) and discontinuous 
for $w>1$, the schematic model shows that this $B$ line cannot terminate at the 
intersection with the $B'$-line. Rather, it continues into the glass region, so that there exist two different 
glass states separated by a line of glass-to-glass transitions. Fig. S6 gives the pertinent states diagram of the model. 
Choosing an overdamped dynamics with initial time-scale $\tau_0$ in both correlators (see Eq. (4.34), p.~203 of 
Ref.~\textit{(41)}), typical correlation functions for the fluid (blue), liquid glass (orange), and glass (green) state are 
depicted in Fig. S6 as well. In liquid glass states, $\Phi_1(t)$ arrests at a finite plateau while $\Phi_2(t)$ decays to 
zero.

\subsection*{References for Supplementary material}

\begin{enumerate}
\item[49.] Happel, J. Brenner, \textit{H. Low Reynolds number hydrodynamics} (Springer Netherlands, 1983).
\item[50.]  Krieger, U. Bosse, J. $\alpha$ relaxation of a simple molten salt near the glass transition. \textit{Phys. Rev. Lett.} \textbf{59}, 1601?1604 (1987).
\item[51.] G\"{o}tze, W. \textit{Complex Dynamics of Glass-Forming Liquids: A Mode-Coupling Theory} (Oxford University Press, 2009).

\end{enumerate}

\end{document}